\documentclass[runningheads]{llncs}
\usepackage{amssymb}
\usepackage{pifont}
\usepackage{subfig}
\usepackage{multirow}
\usepackage[]{siunitx}
\usepackage{hyperref}
\usepackage[all]{hypcap}

\usepackage[table,xcdraw]{xcolor}
\usepackage{graphicx}
\begin{document}
\title{Side-Channel Aware Fuzzing}
\author{Philip Sperl \and Konstantin B{\"o}ttinger}
\authorrunning{P. Sperl \and K. B{\"o}ttinger}
%
\institute{Fraunhofer Institute for Applied and Integrated Security
\email{philip.sperl@aisec.fraunhofer.de} \newline \email{konstantin.boettinger@aisec.fraunhofer.de}}
\maketitle              
\begin{abstract}
Software testing is becoming a critical part of the development cycle of embedded devices, enabling vulnerability detection.
A well-studied approach of software testing is fuzz-testing (fuzzing), during which mutated input is sent to an input-processing software while its behavior is monitored. 
The goal is to identify faulty states in the program, triggered by malformed inputs. 
Even though this technique is widely performed, fuzzing cannot be applied to embedded devices to its full extent.
Due to the lack of adequately powerful I/O capabilities or an operating system the feedback needed for fuzzing cannot be acquired. 

In this paper we present and evaluate a new approach to extract feedback for fuzzing on embedded devices using information the power consumption leaks. 
Side-channel aware fuzzing is a threefold process that is initiated by sending an input to a target device and measuring its power consumption. 
First, we extract features from the power traces of the target device using machine learning algorithms. 
Subsequently, we use the features to reconstruct the code structure of the analyzed firmware. 
In the final step we calculate a score for the input, which is proportional to the code coverage. 

We carry out our proof of concept by fuzzing synthetic software and a light-weight AES implementation running on an ARM Cortex-M4 microcontroller. 
Our results show that the power side-channel carries information relevant for fuzzing. 
\end{abstract}

\section{Introduction}
Embedded systems are nowadays used in a wide range of domains and applications.
The employed devices are often connected to information exchanging networks like the Internet of Things (IoT).
As the number of connected devices is continuously growing, the security of the employed software is gaining impact on our daily life.

Vulnerabilities in embedded devices can be classified by their cause of occurrence, e.g., programming errors, web-based vulnerabilities, weak access control or authentication, and improper use of cryptography \cite{7232966}. 
In the following we will focus on the security threats emerging from programming errors. 
Such errors often lead to memory corruptions like buffer overflows, which attackers make use of in targeted exploits. 
To prevent or find errors leading to such vulnerabilities, several measures exist. 
For instance, source code analysis during the development of the system, or subsequent reverse engineering. 
Both techniques require either the source code or at least deep understanding of the system. 
Furthermore, both approaches cannot be automated or executed large-scale, which increases cost, either during the development or security evaluation phase. 

From the realm of general purpose computers and software testing, the approach of fuzz-testing (fuzzing) is widely accepted and even executed during commercial software development \cite{Howard:2006:SDL:1202957}. 
During fuzzing, automatically generated input is sent to the input-processing software under test (SUT) while its behavior is examined. 
Different instrumentation techniques and mechanisms provided by the operating system (OS) allow evaluation of the impact of the input, leading to an effective and automated vulnerability detection tool. 
Because of the I/O-limitations, restricted computing power, and missing OS, embedded systems lack the possibility of returning enough information required during fuzzing \cite{muench:ndss18}. 
This restrains fuzzing on embedded devices to a black-box manner. 

To circumvent this problem, we present a novel and unexplored technique using the power side-channel of embedded devices as source of feedback for fuzzing. 
We show that the process of sending an input vector to the device, measuring the power consumption, and extracting information from the power traces enables us to deduce the control flow graph (CFG) of the software. 
Subsequently, we present a method of using this representation of the control flow to evaluate the impact of the input on the behavior of the device in terms of code coverage. 
We show that this approach is a significant step towards white-box fuzzing on embedded devices.

In summary, we make the following contributions: 
\begin{itemize}
\item We present the novel idea of extracting feedback from the power side-channel to enhance fuzzing of embedded devices. 
\item In the proof of concept we successfully fuzz synthetic software and a light-weight advanced encryption standard (AES) \cite{Standards2001} implementation and evaluate our results. 
\item Finally, we provide a base line for future work on side-channel-based feature extraction.
\end{itemize}

The rest of this paper is organized as follows. 
In Section \ref{related_work} we review related work and summarize latest findings in embedded device fuzzing.
We present our concept of extracting feedback from the power side-channel to enable white-box fuzzing of embedded devices in Section \ref{concept}.
In Section \ref{experiments} we present our proof-of-concept and evaluation results.
We conclude the paper in Section \ref{Conclusion}.

\section{Related Work} \label{related_work}
We divide this section into four parts.
First we show challenges in embedded device fuzzing before we introduce state-of-the-art techniques.
Subsequently we present latest findings in side-channel based reverse engineering.
Finally, we present the first contributions using side-channel information to assess the program flow of embedded devices.

Muench et al. \cite{muench:ndss18} split the challenges encountered when fuzzing embedded devices into three categories.
We adopt this categorization for devices without an OS and explain it in the following.

Fault Detection: 
Faults during program execution often lead to program crashes, which need to be detected during fuzzing.
For fuzzing on a PC, the OS offers mechanisms for detecting crashes, e.g., a segmentation fault communicated to the fuzzing tool by console output. 
Even though some embedded processors contain Memory Management Units (MMU), the lack of an OS hinders the possibility to communicate faults to the outer world.
Hence, possible program crashes or subsequent reboot sequences may remain undetected.

Performance, Throughput, and Scalability: 
Profiting from the multitasking capabilities of OS-based systems, multiple instances of the SUT can run simultaneously, increasing the fuzzing throughput. 
Transferring this knowledge to embedded systems suffers limitations, because of the missing OS or single-core architecture of the devices. 
A solution is the application of multiple devices executing the same firmware, which may be limited due to financial restrictions. 

Instrumentation: 
Both, compile-time and run-time approaches suffer in feasibility due to the restrictions in embedded systems. 
Recompiling binaries, like on PCs, is not possible if the source code is not available. 
Besides the difficult task of binary rewriting, run-time instrumentation techniques as well as virtualization are not applicable due to the missing OS.

Because of the difficulties presented above, research concerning fuzzing of OS-free embedded devices exclusively deals with a black-box approach.
Even though we present a white-box solution, some findings in black-box fuzzing are worth mentioning. 
Koscher et al. \cite{Koscher:2010:ESA:1849417.1849990} and Lee et al. \cite{7098059} carry out fuzzing of CAN packets.
Alimi et al.\cite{6903734} made use of black-box fuzzing when looking for vulnerabilities in applets on smart cards. 
Muench et al.\cite{muench:ndss18} try to improve fuzzing by partial and full emulation of embedded devices. 
Zaddach et al. \cite{EURECOM+4158} propose a related approach and improve the emulation-based testing process by forwarding memory accesses on the emulated target to the physical device.

Symbolic execution poses an alternative approach. 
Davidson et al. \cite{182944} show the possibility to find bugs in firmware using a specification of the memory layout of the target device, the source code, and their KLEE-based \cite{Cadar:2008:KUA:1855741.1855756} symbolic execution engine. 
With this setup the authors are able to perform fuzzing of the target device. 
Further improvements to this approach were shown by Corteggiani et al. \cite{217620}, in which access to the source code is required as well.

Ever since the introduction of the Differential Power Analysis by Kocher et al. \cite{Kocher:1999:DPA:646764.703989} numerous publications picked up the concept of analyzing information leaked over the power or electromagnetic (EM) side-channel of embedded devices.
As we link the concepts of side-channel analysis and fuzzing, we profit from research in the field of side-channel based reverse engineering. 
Strobel et al. \cite{7092372} show an effective way to reconstruct the assembly code executed by CPUs using EM emanations. 
By decapsulating the attacked chip and using eight measurement probes, the authors achieve an instruction recognition rate of \num{87,69}$\%$ targeting real code. 
Msgna et al. \cite{Msgna:2014:PIS:2724797.2724811} use a k-nearest neighbors (\textit{kNN}) classification for reverse engineering. 
The authors exploit the fact that the power consumption of digital circuits depends on the processed data and executed instructions \cite{Mangard:2007:PAA:1208234}. 
Targeting test code the authors achieve an instruction recognition rate of \num{100}$\%$.

The first method using side-channel information to gain insights into the executed code paths of an embedded device is presented by Callan et al. \cite{Callan:2016:ZPV:2931037.2931065}. 
In the training phase the authors measure the EM emanations of a target device while it executes instrumented code.
During the profiling phase EM traces are measured while the device executes the original source code. 
The authors compare both sets of EM traces to identify the currently executed program path.
If further refined, this approach poses a possible source of information to fuzz firmware for which the source code is available.
Nonetheless, the authors do not evaluate the scenario in which the source code is not available.
Similarly, Han et al. \cite{han2017watch} use the EM emanations of embedded controllers to identify the current control flow.
The authors are able to detect malicious executions with high confidence.
Van Aubel et al. \cite{10.1007/978-3-319-99843-5_19} use EM traces and methods from the classical side-channel analysis to build an intrusion detection tool for industrial control systems.

\section{Elements of Side-Channel Aware Fuzzing} \label{concept}
In this section we present our main contribution, a novel technique for extracting feedback from an embedded device using the power side-channel to make white-box fuzzing possible. 
In particular, we calculate scores for the inputs proportional to the provoked code coverage. 
The calculation of the scores is inspired by the procedure implemented in the widely used and accepted American Fuzzy Lop (AFL) fuzzing tool. 
By instrumenting the code, AFL counts the basic block transitions during the processing of each input. 
The number of newly executed basic block transitions is then directly used as score for the inputs, see \cite{AFL}. 
In subsequent fuzzing runs, AFL mutates inputs with the highest scores and feeds them back to the SUT.
The goal is to accumulate a series of inputs for which a code coverage of \num{100}$\%$ is reached.
In this paper we neglect the prioritization and mutation of the inputs and solely provide code coverage scores.
The scores can then be fed to a tool like AFL, to perform the remaining actions required for a closed fuzzing loop.

\begin{figure}[]
\centering
\includegraphics[width=8.5cm]{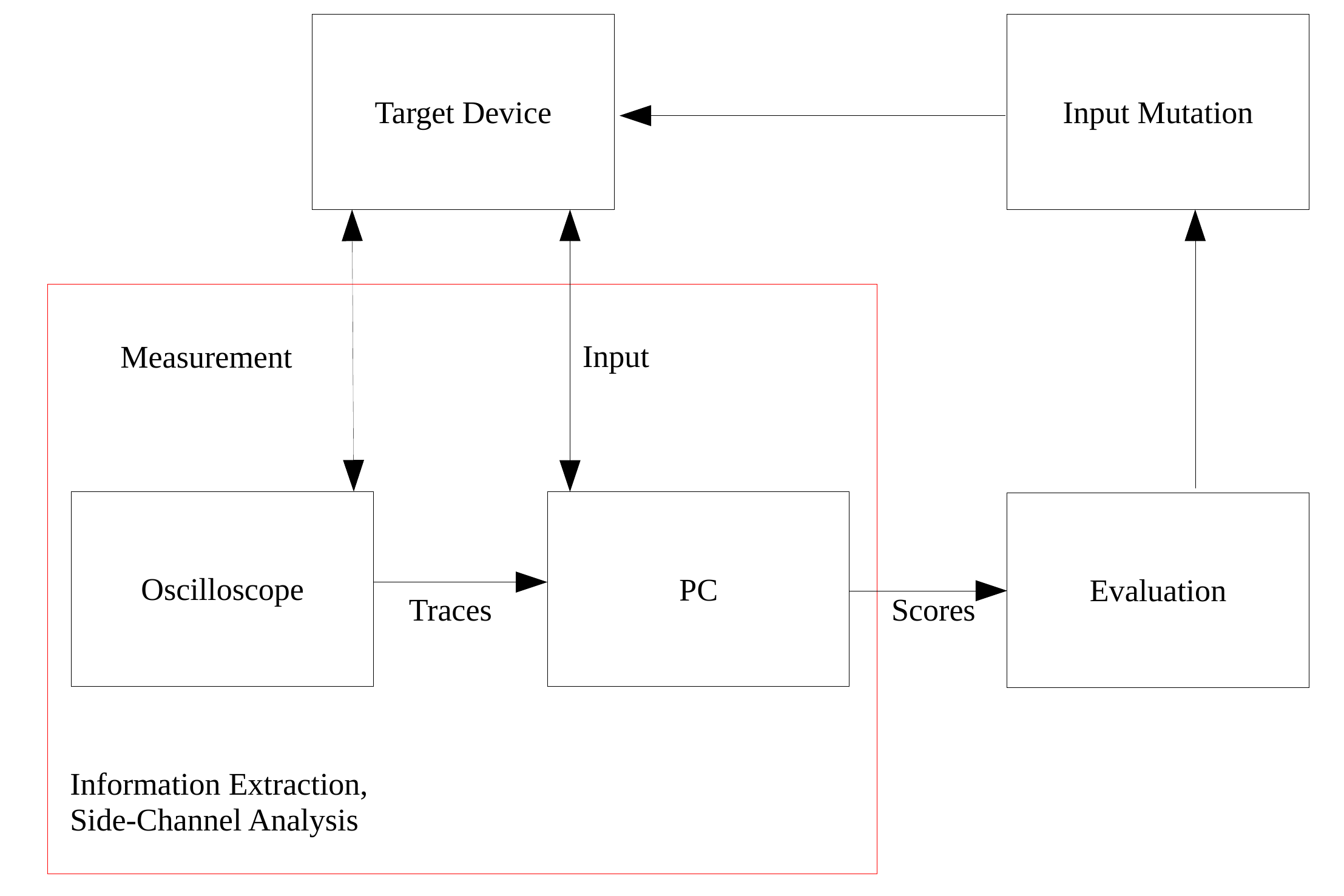}
\caption{Side-channel driven fuzzing feedback loop.}
\label{complete_setup}
\end{figure}

We carry out the illustration of our concept in a bottom-up manner. 
First, we explain all underlying building blocks, before we present the complete concept as well as the overall schematic of the side-channel aware fuzzing loop.
In Figure \ref{complete_setup} we show the setup and required equipment. 
During the feedback-driven loop, we use an oscilloscope to measure the power consumption of the target device and send the power traces to the evaluation computer. 
Using this PC we process the traces and conduct the side-channel analysis (SCA) consisting of three steps.
First, we identify the individual basic blocks of the software. 
In the second step we characterize the found basic blocks and the transition sequence. 
Finally, we calculate code coverage scores for each input.

Throughout the discussion of the SCA building blocks we assume that for each input sent to the device, its power consumption is measured. 
The lengths of the power traces cover the exact time the device requires to process the currently evaluated input.
For the sake of simplicity, we examine one input and its corresponding trace in the following.
In a real world application and during our proof of concept, we perform the analysis for each input sent to the target device.

\subsection{Feature Extraction Using the Power Side-Channel} \label{why_branches} 
A power trace is an array of \textit{n} quantized measurement points of the physical power consumption of the target device. 
The number of points depends on the measurement period and the sampling rate of the used oscilloscope.
Even though side-channel aware fuzzing is based on power traces, we convey our concept without illustration of actual power measurements.
We decided to do so in order to provide a general introduction of our approach, which can be transferred to a wide range of devices.
The central goal of our concept is to estimate the code coverage triggered during the processing of the current input.
A widely used metric to express code coverage is the number of basic block transitions.
In this paper we make use of this idea, therefore we define basic blocks and the separation of such in the following.

Basic blocks are lines of code which are executed without interruption of a branch instruction \cite{Hennessy:2011:CAF:1999263}. 
Hence, if the execution of a basic block begins, it will be executed completely before another basic block is triggered. 
Branch instructions at the end of each basic block coordinate the transitions. 

This observation builds the core of our concept.
If we detect the moment in which a target device executes a branch instruction, we find the borders between the basic blocks.
As a consequence, we focus on this class of instructions and present a method to detect branch instructions using the collected power traces.
Subsequently we provide methods to characterize the basic blocks, so that we can estimate the number of individual basic block transitions per input.

\subsubsection{Branch Detection} \label{Branch_Detection}
The power consumption of digital circuits consists of four components.
Each sample of a power trace is determined by the executed instruction, processed data, electronic noise characteristics of the device, and a constant component.
To detect branches we exploit the operation-dependent component of the power consumption.

We interpret the branch detection as a binary classification problem, since all remaining instructions executed by the device are not of interest. 
During initialization we split the examined power trace into \(k\) windows. 
Note that the beginning of each window matches the beginning of a potential branch instruction. 
Moreover, the windows and the branch instructions share the same length.
After this initialization, we carry out the binary classification, which we execute \(k\) times per trace. 
In this classification procedure a predictor decides which window shows a branch instruction. 
Machine learning algorithms with a previous training phase can build the basis for the predictor. 
If the prediction indicates that the currently analyzed window shows a branch, we add the location to the result list \(B_{\text{locations}}\). 

To create train data on which the machine learning algorithm can be trained, we need an identical and programmable target device.
We let this training device execute branch instructions with various offsets and distances, while we measure its power consumption.
During the supervised training, we use the labeled power traces to create the branch detection model.

\subsubsection{Branch Distance Classification} \label{Branch_Distance} 
A feature providing evidence whether a transition was already executed before is the distance between a found basic block and its predecessor.
Here, we define this distance as the number of instructions which the device under test (DUT) skips due to the branch instruction.

In order to estimate the number of skipped instructions we exploit the data-dependent component of power traces and the following insight.
Branch instructions contain labels to which they should jump. 
The CPU calculates the distance between the current location and the label and adds it to the program counter. 
Since this process is data-dependent, the power side-channel leaks information about this distance. 

We again use a supervised machine learning algorithm to estimate the distances of the found branches.
In the training phase our train device performs branches with known distances while we measure its power consumption.
We use the resulting labeled data to train the algorithm and create the branch distance model.

In the branch distance classification we first cut the traces, such that we only evaluate samples measured during the execution of branch instructions.
We apply our previously created model and store the classification output in the result list \(B_\text{{distances}}\).

As an alternative to the branch distance classification which might not be successful for every branch or device, we present an additional approach to distinguish the basic blocks in the following.

\subsubsection{Basic Block Fingerprinting} \label{fingerprints}
To distinguish basic blocks we assign side-channel based fingerprints to each. 
For this purpose we use the slices of the power traces between the previously found branch instructions.
These parts represent the power consumption during the execution of basic blocks.
Hence, we conduct an initialization phase in which we cut the traces accordingly.
In the first step of our algorithm we extract four features from each analyzed power trace window.
Subsequently we use the features to fingerprint the basic blocks. 
We present the extracted features and illustrate the purpose of each in terms of contributed information.

Basic blocks often differ in their required execution time. 
Evaluating this feature allows an easy-to-implement distinction. 
Therefore, the first metric we consider is the length \(P_\text{{length}}\) of the individual slices. 

For the calculation of the second feature \(P_\text{{peaks}}\), the algorithm evaluates the number of peaks for each trace segment. 
For the majority of embedded devices, the power traces consist of periodically occurring peaks. Internal clocks which may have the frequency of the system clock or other clocks like the flash clock have the major impact on the number of peaks. 
Additional peaks can occur in the power traces due to complex instructions. 
The additionally required computational power increases the power consumption resulting in spikes in the traces. 
Thus, the number of peaks in the trace windows shows the approximate duration and indicate the complexity of the executed instructions. 

The third metric is the mean \(P_\text{{mean}}\), of the windows, which is the mean power consumption during the basic block execution. 

The last metric we calculate is the skewness, \(P_\text{{skewness}}\). 
Translated to the power consumption, this metric enables the following distinction of cases. 
Assume two basic blocks \(A\) and \(B\) sharing the same number of instructions, with identical mean power consumption. 
Basic block \(A\) consists of instructions with evenly distributed computational cost and resulting power consumption. 
In contrast to that, basic block \(B\) contains one significantly more complex instruction than the instructions found in \(A\). 
Furthermore, the remaining instructions in \(B\) are less complex than the ones found in \(A\), resulting in a the same mean power consumption. 
By analyzing the skewness of the power traces, we are able to distinguish basic blocks \(A\) and \(B\).

In the final step of the basic block fingerprinting, we superpose the four extracted features to create the fingerprints. 
We present two approaches to achieve this superposition. 

In our first approach we take all four values of the currently analyzed basic block and store them in a four-dimensional vector. 
We call this method \textit{separated}.
This approach is easy to implement, however, in subsequent calculations the dimension of the feature vector can lead to increased execution times compared to a scalar value representing the fingerprint. 

Therefore, in the second approach, which we call \textit{summed}, we adopt this idea and sum up the four previously calculated values. 
Alternatively, a hash function can be applied to generate a fingerprint. 
This approach is very effective if the underlying values already lead to a strong distinction between the basic blocks. 
For both approaches we store the fingerprints of the basic blocks in the result table \(B_\text{{prints}}\).

\subsection{Control Flow Reconstruction} 
\begin{figure}[]
   \centering
   \subfloat[][\centering CFG-RI.]{\includegraphics[width=.22\textwidth]{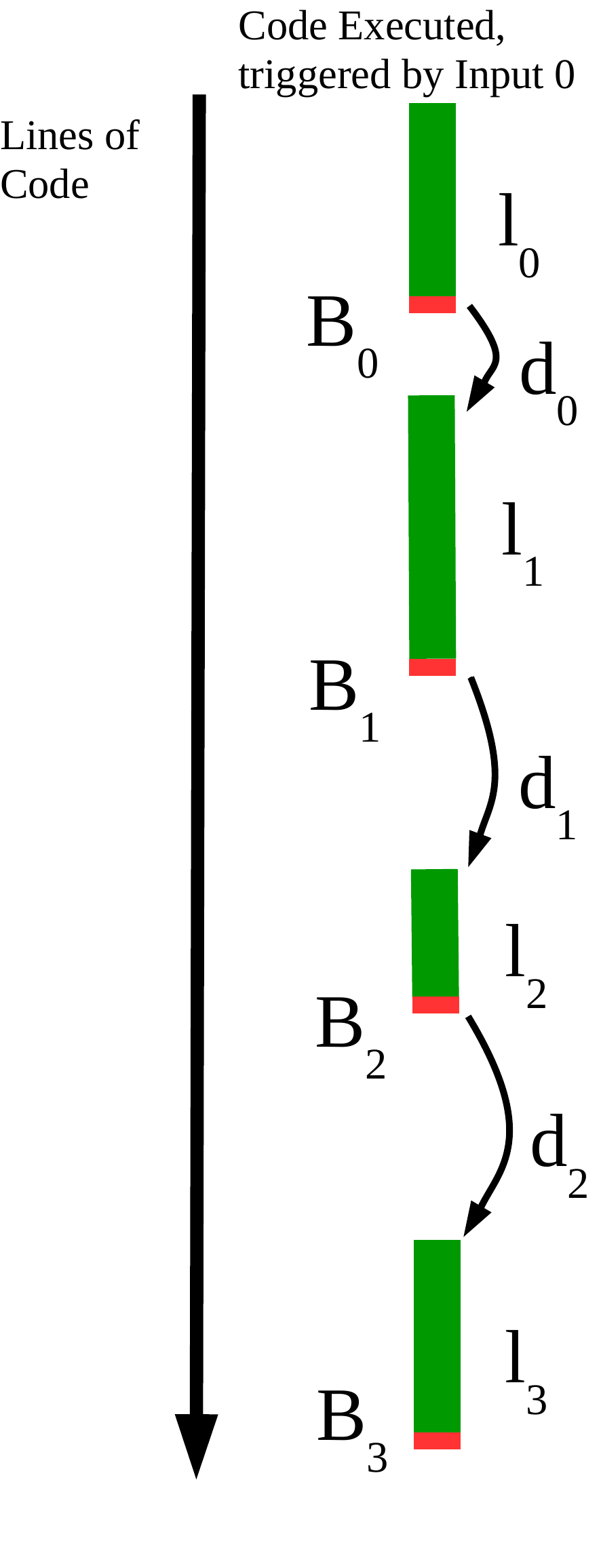}} \hspace{24mm}
   \subfloat[][\centering CFG-RII.]{\includegraphics[width=.22\textwidth]{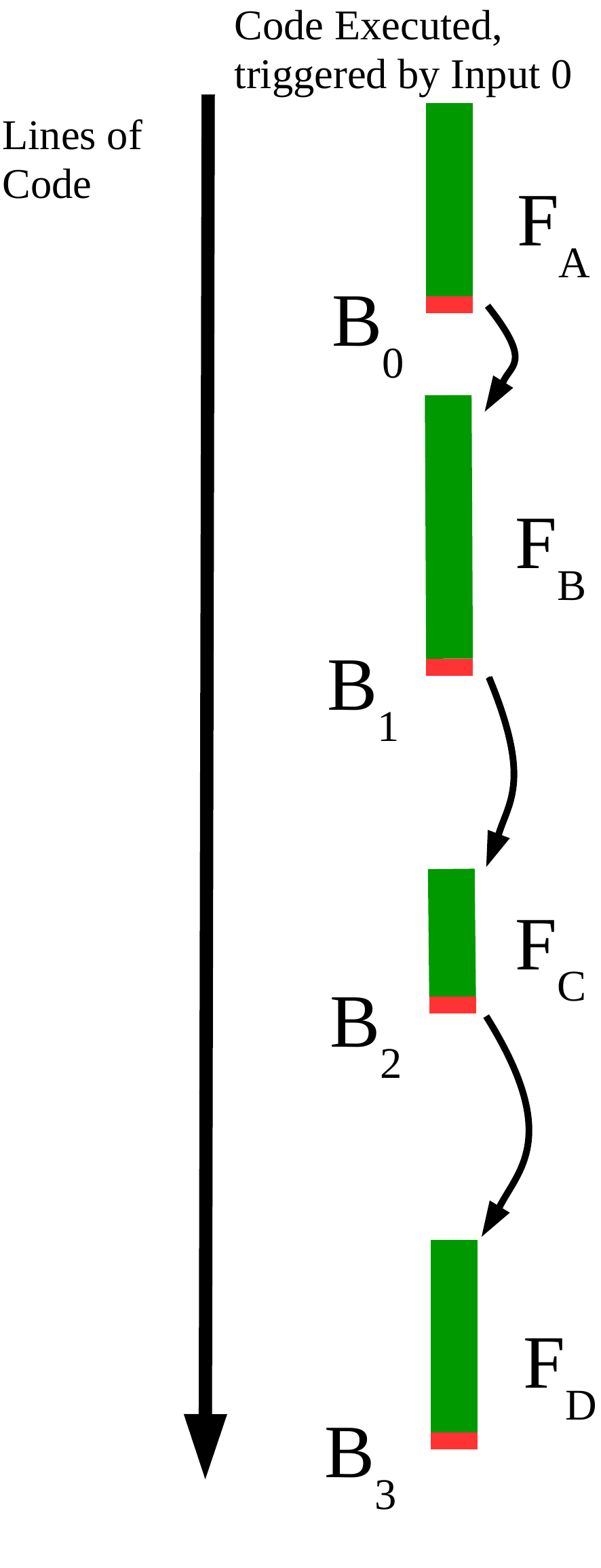}}
   \caption{Illustration of the two control flow reconstruction algorithms and the required information.}
   \label{cfg_recon}
\end{figure}

In this section we use our knowledge of the basic block transitions to reconstruct the program flow of the analyzed firmware.
We present two algorithms for this purpose. 
Both use the previously found branch locations. 
The first algorithm uses the branch distances, while the second one uses the basic block fingerprints to further characterize the transitions. 
For both approaches we give an exemplary control flow in Figure \ref{cfg_recon} to visualize the concepts of the reconstruction. 
Each control flow represents the processing of one input by the target device. 

Furthermore, we introduce the following notation. \(B_{x}\) indicates the branch location with the index \(x\). The corresponding distance of the branch to its successor is \(d_x\). The length of the executed basic block with index \(i\) is \(l_{i}\). The fingerprint for the basic block \(A\) is \(F_A\). 

We store the results of the control flow reconstruction in the table \(T_\text{{CFG}}\). Table \ref{cfg_table} shows the results according to the examples from Figure \ref{cfg_recon}. The left column holds the branch IDs for all inputs, while the columns to the right hold the results of the control flow reconstruction. 
In the following we convey both algorithms and explain the results from Table \ref{cfg_table}.

\subsubsection{CFG Reconstruction I (CFG-RI)}
In this CFG reconstruction algorithm we use our knowledge of the branch locations \(B_\text{{locations}}\) and their corresponding distances \(B_\text{{distances}}\).
For each branch we calculate a two-dimensional vector which characterizes the subsequent basic block and store it in \(T_\text{{CFG}}\).
Thereby we sufficiently describe the control flow of the tested software in order to evaluate the sequence of performed basic block transitions. 
In the following we describe the creation of the vectors using Figure \ref{cfg_recon} a).

We build the vector \([B_\text{{offset}}, C_\text{{length}}]\) by stacking two characteristic values for each basic block. 
\(B_\text{{offset}}\) indicates the offset of the branch location in the code, expressed in number of instructions executed or skipped until the branch itself is executed. 
In Figure \ref{cfg_recon} a), for branch \(B_0\) this value is \(l_0\). 
For branch \(B_1\) this value depends on the following intermediate results. 
Until \(B_1\) is executed, the two basic blocks with the lengths \(l_0\) and \(l_1\) are executed.
In addition to that, we estimate the distance of branch \(B_0\) to be \(d_0\). 
Hence, with the sum of the discovered distances (\(l_0\) \(+\) \(l_1\) \(+\) \(d_0\)) we express the offset for branch \(B_1\). 
The second result value \(C_\text{{length}}\) is the length of the code after each branch, \(l_i\).
For the branches \(B_0\) and \(B_1\) these values are \(l_1\) and \(l_2\), respectively. 

With the location of a specific branch with respect to the previously executed code and the length of the following basic block, we sufficiently describe the control flow of the code to evaluate if we triggered a new basic block transition.

\subsubsection{CFG Reconstruction II (CFG-RII)} 
In this approach we use the unique fingerprints \(B_\text{{prints}}\). 
Each fingerprint describes one basic block, which enables us to distinguish them.
Hence, we directly store the fingerprints in the result table next to the according branch ID. 
Note that we assume the code started without an initial branch. 
Therefore the first fingerprint is stored without an associated branch ID. 
With this result table we provide the sequence of transitions and fully reconstruct the program flow on the basic block level.

\begin{table}[]
\caption{Result table \(T_{CFG}\) showing the results for both control flow reconstruction algorithms.}
\centering
\begin{tabular}{l|l|l}
\textbf{Branch ID}    & \textbf{Results CFG-RI} & \textbf{Results CFG-RII}     \\ \hline
\begin{tabular}[c]{@{}l@{}}-\\ 0\\ 1\\ $\vdots$\end{tabular}      &   \begin{tabular}[c]{@{}l@{}}\([B_{offset,start}, C_{length,start}]\)\\ \([B_{offset,0}, C_{length,0}]\)\\ \([B_{offset,1}, C_{length,1}]\) \\ $\vdots$\end{tabular}       & \begin{tabular}[c]{@{}l@{}}\(B_{prints, start}\)\\ \(B_{prints, 0}\)\\ \(B_{prints, 1}\)\\$\vdots$\end{tabular}       \\ \hline
$\vdots$      &   $\ddots$    &   $\ddots$
\end{tabular}
\label{cfg_table}
\end{table}

\subsubsection{Theoretical Comparison} 
Both algorithms reconstruct the analyzed software to such an extent, that we are able to calculate scores for each input representing the number of newly triggered basic block transitions.
The two approaches differ in the used side-channel information.
Hence, each algorithm has advantages depending on the attacked device and measurement quality. 
In the following we give a recommendation on when to use either of the algorithms. 

CFG-RI can lead to an accurate reconstruction of the examined code. 
The drawback of this approach is the additional training phase we need to perform prior to the branch distance classification. 
This leads to a more time and memory consuming process before the actual fuzzing. 
In order to classify the branch distances, the corresponding machine learning model has to be loaded in the evaluation computer in addition to the model for the branch detection and the analyzed power traces. 
Moreover, the measurement quality and target device properties highly influence the accuracy during the estimation of the branch distances. 
Different test devices and measurement equipment may lead to poor results preventing a correct estimation. 
Additionally, we emphasize the fault propagation concerning this algorithm. 
If one branch distance is classified wrongly, the remaining code reconstruction process results in a flawed CFG. 

In CFG-RII the results during the fingerprint calculations do not depend on the quality of the measurement setup and the attacked device as it is the case during the branch distance classification. 
Furthermore, a potential error does not corrupt all following results. 
As drawback, regardless of the complexity of the fingerprints, the probability of collisions is not fully ruled out. 
A collision occurs if for two or more different basic blocks the same fingerprint is generated.
The consequence of such an error would be in the worst case, that one yet unknown basic block transition would not be detected as such. 
The resulting score of the analyzed input would be smaller than the actual score.

We sum up the findings of the comparison as follows.
With algorithm CFG-RI we can precisely reconstruct the structure of the tested firmware.
Because of the error propagation property, we exclusively recommend using it if a strong recognition rate during the branch distance classification is reached.
In contrast to that, we present CFG-RII as an easy to implement and intuitive backup strategy.
We will present a quantitative comparison of both algorithms in Section \ref{experiments}.

\subsection{Score Calculation}
In the final step of our approach we calculate the score which is the number of newly triggered basic block transitions per input. 
The list $\Omega$, that is empty at the beginning of the fuzzing process, holds all known basic block transitions. 
For the score calculation we use the result table we gained during the control flow reconstruction.
We analyze the neighboring pairs of basic blocks and their corresponding representation, realized either with the fingerprints or the branch distances.
If the currently analyzed pair is already stored in $\Omega$, we will not increase the score. 
In contrast to that, if the pair is not in $\Omega$, we add it to the list and increase the score for the corresponding input by \num{1}.
With this procedure we adopt the concept of estimating the code coverage in terms of basic block transitions, similar to the AFL fuzzing tool.

\subsection{Error Prevention and Trace Preprocessing} \label{errors}
To prevent errors, we aim to increase the signal-to-noise-ratio (SNR) of the measured power consumption. 
Since the electronic noise follows a normal distribution \cite{Mangard:2007:PAA:1208234}, a widely performed approach is to increase the number of measurements showing the same operations and form superposed traces.
We adopt this concept and present three different approaches to achieve this. 
For all approaches, we send the same input to the device multiple times and capture the power traces.
Note that the traces need to be aligned correctly in order to allow valid calculations.
For this purpose we use a precise trigger, which depends on the system clock of the DUT to start the measurement.

In the first approach, for every sample point in the power traces, we calculate the average to form a \textit{mean} trace. 
Alternatively, we continuously average over the measured samples and assign higher weights to later recorded traces.
We call this approach \textit{sweep}. 
Using either of the superposed traces we carry out all calculations as explained in Section \ref{concept} resulting in one score per input. 

Alternatively, in the second approach we execute the feature extraction and score calculation for every trace showing the same operations separately. 
After a following majority vote we accept the most probable results. 

In addition, a hybrid version poses a third alternative.
Here we calculate multiple scores for the same input using either \textit{mean} or \textit{sweep} traces.

\subsection{Overall Side-Channel Driven Fuzzing Algorithm}
Above we described all building blocks of our approach.
In this section we link them and present an overview.
Assume we sent multiple inputs to the DUT while we recorded the power consumption during the processing of each input.

In the first step, we load a batch of power traces and calculate the pairwise mean-squared-error (MSE) among them.
With this measure we perform a first refinement prior to the actual SCA calculations to exclude multiple traces showing the same sequence of instructions.
If for a pair of traces, the MSE is below a certain threshold, we can assume the same sequence of operations and hence basic blocks were triggered.
We exclude such traces and increase the overall fuzzing throughput. 
Note that we need to analyze the noise properties of the tested DUT to define the MSE threshold.

\begin{figure}[ht]
\centering
\includegraphics[width=8cm]{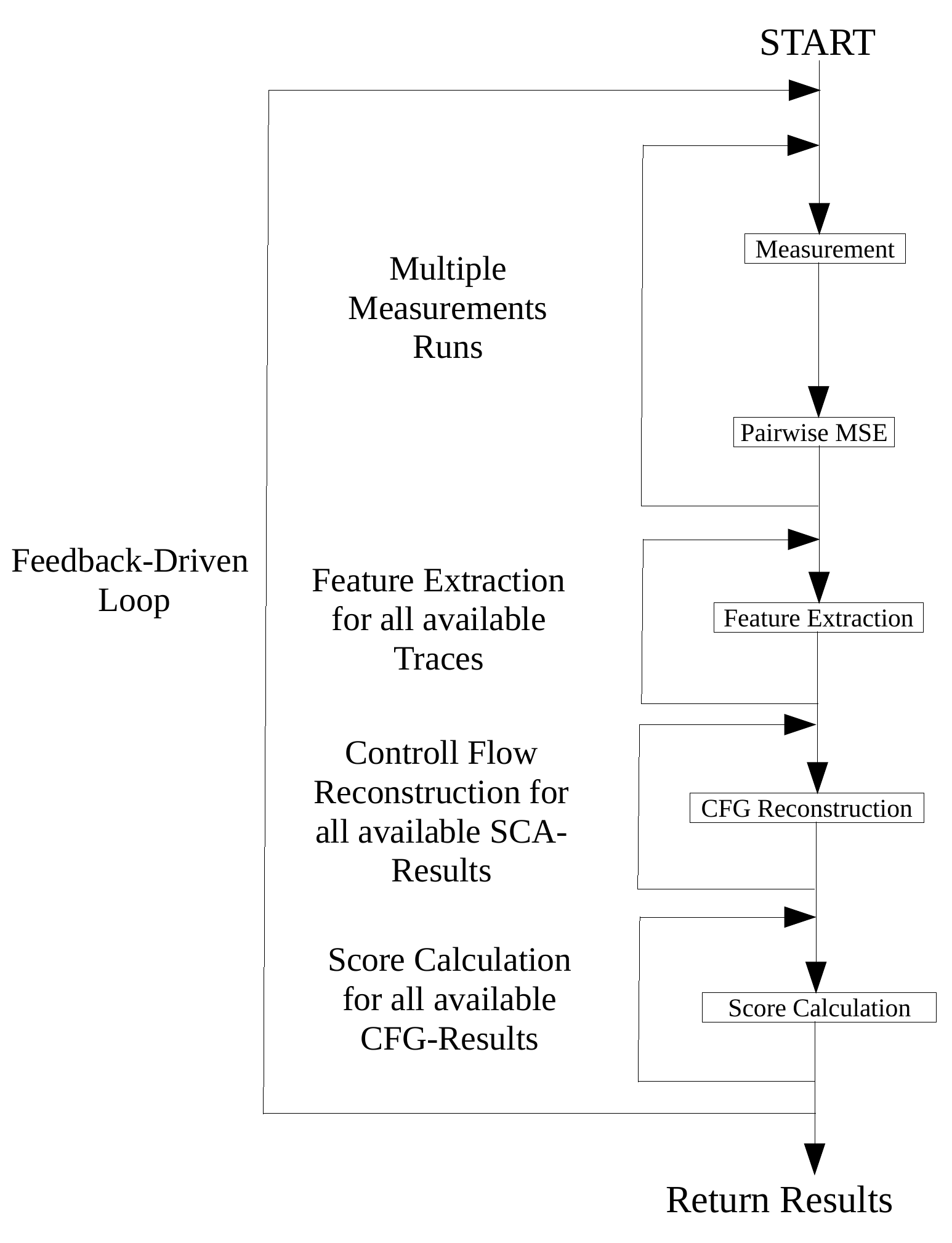}
\caption{Overview of the building blocks of side-channel aware fuzzing.}
\label{embedded_fuzzer_cfg}
\end{figure}

For the remaining traces, we calculate the scores using our previously introduced algorithms. 
To complete the fuzzing loop, we suggest using a state-of-the-art fuzzing tool like AFL or SAGE \cite{godefroid08ndss}. 
The analyst can feed the calculated scores to a tool, which prioritizes the inputs and further mutates promising examples.
Figure \ref{embedded_fuzzer_cfg} shows the overall setup. 
The different loops indicate operations which we execute in parallel to further increase the throughput of the framework.

\begin{figure}[ht]
   \centering
   \subfloat[][\centering No preprocessing.]{\includegraphics[width=.42\textwidth]{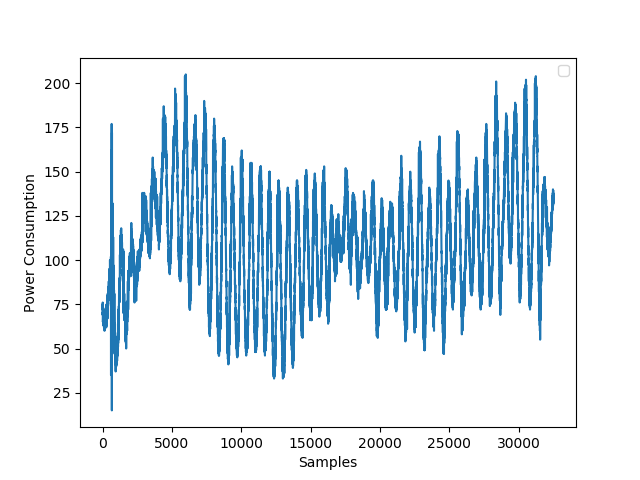}}
   \subfloat[][\centering Mean calculation with \num{10} traces.]{\includegraphics[width=.42\textwidth]{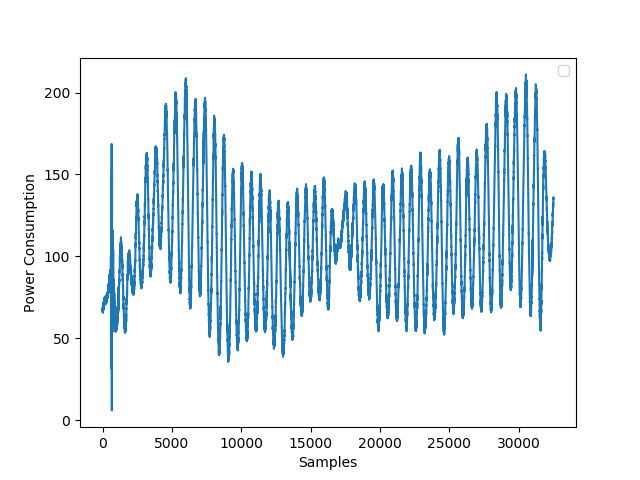}}\\
   \subfloat[][\centering Continuous averaging with \num{10} traces.]{\includegraphics[width=.42\textwidth]{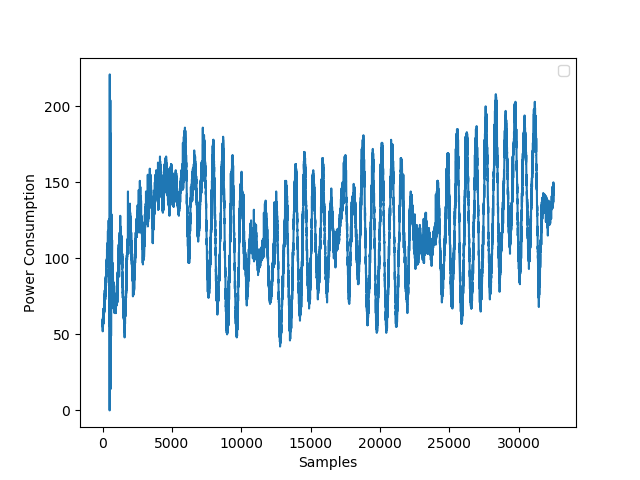}}
   \subfloat[][\centering Mean calculation and continuous averaging, both with \num{10} traces.]{\includegraphics[width=.42\textwidth]{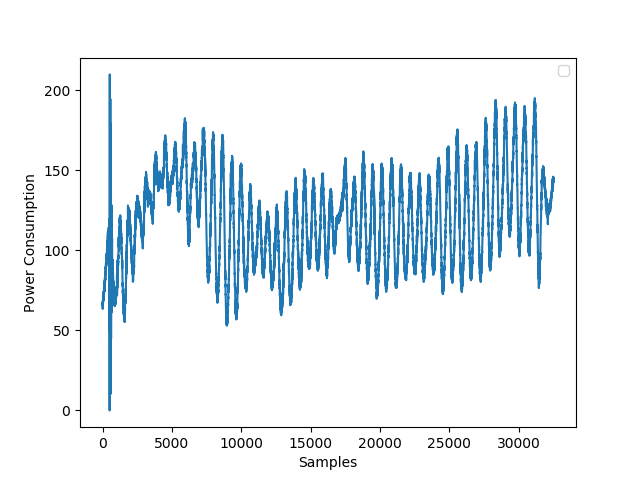}}
   \caption{Four power traces measured during the execution of the same instructions, showing the impact of different preprocessing techniques.}
   \label{trace_preprocessing}
\end{figure}

\section{Experiments and Evaluation} \label{experiments}
The lack of a base line to which different fuzzing techniques can be compared to is a known problem in software testing picked up by Dolan-Gavitt et. al \cite{Dolan-GavittHKL16}. 
The authors present a system, which injects bugs into C code. 
This modified code enables the comparison of various testing tools.

Regarding embedded device fuzzing, we face a similar but more fundamental challenge. 
Prior to evaluating the actual fuzzing success, we need to assess the underlying feedback itself. 
Therefore, we present a test environment, which allows the calculation of theoretical scores. 
Furthermore, we introduce three evaluation metrics to assess the quality of the calculated scores.
This framework forms a possible base line for future work on feedback extraction, aiming to enhance embedded device fuzzing.

Additionally we show implementation details essential for our proof of concept and information about the achieved results, gained during fuzzing of synthetic software and a light-weight AES implementation. 
Finally we discuss the transferability of side-channel aware fuzzing to a broader range of embedded devices.

\subsection{Evaluation Code} \label{test_code} 
During the proof of concept the analyzed target device executes a test software, which allows the calculation of theoretical code coverage scores for the inputs. 
The test software takes \num{16} bytes as input so that \num{16} binary decision stages are passed throughout the processing. 
We test five versions of the software, which differ in the decision probability at the binary decision stages.
Each version covers \num{48} basic blocks and \num{60} possible basic block transitions. 
The evaluation code hence provides \num{300} different basic block transitions, our implementation of side-channel aware fuzzing needs to detect. 
The distances of the branches between the basic blocks range from \num{10} to \num{150} skipped instructions.
For the individual basic blocks we chose a randomized implementation such that they differ in length, instruction types, used operands, and complexity. 
Since embedded devices often operate in sensing-actuating loops, some basic blocks in real-world code may only contain a small and simple sequence of instructions.  
Therefore, to emulate this case the shortest basic blocks contain \num{10} instructions, whereas we use \num{110} instructions to form the longest blocks. 
A further motivation to include short basic blocks is to show the sufficient detail among the power traces which we are able to exploit. 
If we are able to separate even small basic blocks, our concept does not lack applicability regarding real software.

\subsection{Benchmark Metrics}
We consider three metrics as measures for the performance of our concept. 
To allow the following evaluation we interpret multiple scores as arrays or result-traces, respectively. 
This holds true for the calculated as well as for the theoretical scores. 
We evaluate the MSE and correlation coefficient between the result-traces. 
In addition we calculate the number of crucial errors. 
Such errors occur if for a certain input which triggers the execution of at least one new basic block transition, a score of \num{0} is returned. 
In this case the according input would be discarded and not be considered for further mutations.
This error type leads to a major decrease in the fuzzing success.

\subsection{Implementation Decisions - Classification Approach}
The main decision during the implementation is the choice of a machine learning-based classifier for the branch detection and branch distance classification. 
To perform a comparison of different classification algorithms we record \num{50000} traces for the two classes (branch vs. no branch) and use them to train the machine learning models.
We compare eight different machine learning algorithms to detect branches.
To view the performance of each algorithm, we calculate Matthew$'$s correlation coefficient (MCC) \cite{MATTHEWS1975442}, which is well suited for the evaluation of binary classifications of imbalanced data sets \cite{Powers_David_what}. 

We achieve best results using a \textit{kNN} classification with \(k=3\) and an MCC of \num{0.93}. 
This result corresponds to the latest findings in related work, see Section \ref{related_work}.
Therefore, we apply this approach during the proof of concept in which we were able to reach an MCC of \num{0.78} for the branch detection.

\subsection{Test Scenario and Power Traces}
In this section we outline important facts about our proof of concept. 
We show power traces to illustrate the actual application of the algorithms from Section \ref{concept}. 
We particularly focus on the branch detection and branch distance classification.

During the evaluation phase, our test software (see Section \ref{test_code}) runs on an STM32F417 \cite{STM32} microcontroller.
This reduced instruction set computer (RISC) based controller uses the ARM Cortex-M4 \cite{cortex_M4} processor.  
We set the clock frequency of the DUT to \num{84}MHz.
The DUT processes a batch of \num{100} random inputs while we measure its power consumption using a shunt resistor of \num{47}$\Omega$.
From the Nyquist-Shannon sampling theorem \cite{Shannon1949} we know that the sampling rate needs to be at least twice as high as the frequency of the measured signal to prevent a loss of information.
Hence, we set the sampling rate of our \textit{LeCroy WavePro 760Zi-A} oscilloscope to \num{5}GS/s.
After all steps of side-channel aware fuzzing our implementation returns a result-trace containing \num{100} scores.
Each score corresponds to one input sent to the DUT.
Figure \ref{trace_preprocessing} shows four power traces we measured with a differential probe. 
The individual traces differ in the applied preprocessing technique and give the reader an intuition about the form of the analyzed data. 

\begin{figure}[ht]
\centering
\includegraphics[width=8cm]{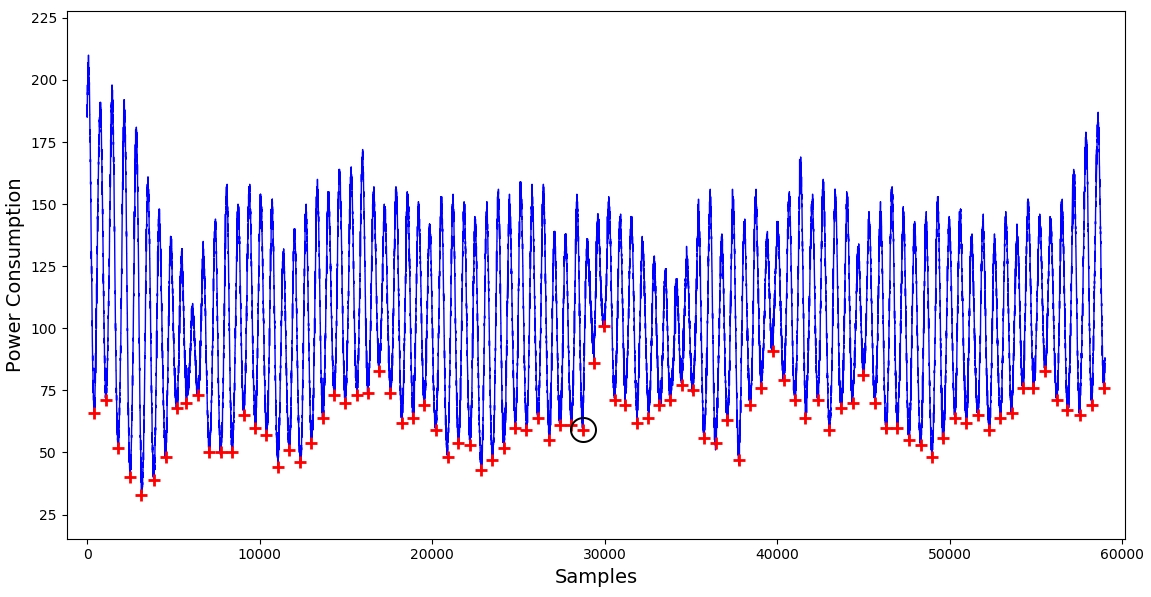}
\caption{Power trace during the branch detection, consisting of a peak detection and binary classifications.}
\label{branch_detection}
\end{figure}

Figure \ref{branch_detection} shows a power trace during the branch detection. 
For a simple illustration, we chose to implement one branch instruction in the code executed by the DUT. 
In the first step we slice the power trace into equally sized windows, using a peak detection. 
Each peak, marked with a red cross is a potential beginning of a branch instruction. 
The windows have the length of one branch instruction, which we characterized in the training phase. 
After the binary classification on each window, only one peak is detected as the beginning of a branch, marked with a black circle. 

\begin{figure}[]
\centering
\includegraphics[width=7.5cm]{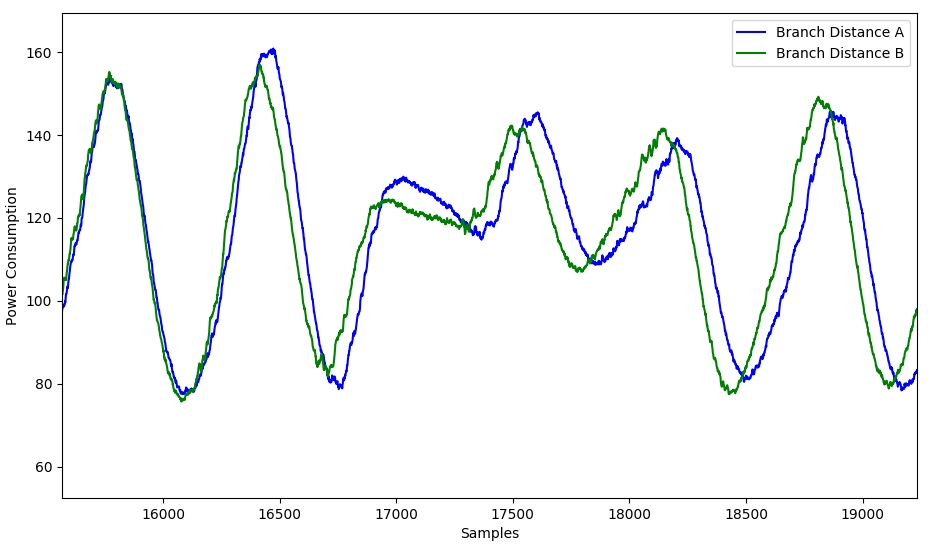}
\caption{Two power traces during the execution of two branches with different distances.}
\label{2_distances}
\end{figure}

Figure \ref{2_distances} shows two power traces during the branch distance classification.
The traces show the power consumption of the target device during the execution of two branches with different distances. 
We can clearly distinguish the power traces and hence the distances of the branches.
Note that both traces still show a large similarity, such that we are able to detect both branches.

\subsection{Results}
We sum up the results of our proof of concept in Table \ref{overall_results}. 
The result values show the effects of the different preprocessing techniques, the two CFG reconstruction approaches, and impacts of a majority vote in the feature extraction. 
We emphasize that preprocessing is a necessary step as we achieved better results compared to an analysis using unprocessed traces. 
The various CFG reconstruction algorithms perform comparably, this allows a selection depending on the analyzed device and measurement quality. 
Regarding our test environment, no majority vote in the feature extraction reaches superior results compared to a majority vote-based approach.

With our results we can report that the power side-channel of embedded devices carries information relevant for fuzzing.
We are able to calculate result traces, which strongly correlate with the actual result values. 
The maximum correlation coefficient is \num{0.95}.
Furthermore, our framework makes on average \num{0.69} crucial errors per \num{100} inputs.

\begin{table}[]
\caption{Results during the proof of concept. The target devices executes synthetic code and handles \num{100} random inputs.}
\centering
\begin{tabular}{l|l|l|ll|ll|ll|}
                                                    & \multicolumn{2}{l|}{\textbf{\begin{tabular}[c]{@{}l@{}}Preprocessing, \\ number of \\ used traces\end{tabular}}} & \multicolumn{6}{l|}{\textbf{\begin{tabular}[c]{@{}l@{}}CFG Reconstruction\\ majority (white) vs.\\ non-majority vote (gray)\end{tabular}}}                                                                                                              \\ \cline{2-9} 
\multirow{-2}{*}{\textbf{Performance Metrics}}      & \textit{\textbf{mean\hspace{0.27cm}}}                                 & \textit{\textbf{sweep}}                                 & \multicolumn{2}{l|}{\textit{\textbf{CFG-RI}}}& \multicolumn{2}{l|}{\textit{\textbf{\begin{tabular}[c]{@{}l@{}}CFG-RII\\ (separate)\end{tabular}}}} & \multicolumn{2}{l|}{\textit{\textbf{\begin{tabular}[c]{@{}l@{}}CFG-RII\\ (summed)\end{tabular}}}} \\ \hline
                                                    & 1                                                      & 1                                                       & 6.16 \hspace{0.05cm}     & \cellcolor[HTML]{EFEFEF}3.21      & 6.73   \hspace{0.05cm}                               & \cellcolor[HTML]{EFEFEF}4.13                                & 5.96     \hspace{0.05cm}                           & \cellcolor[HTML]{EFEFEF}5.23                                \\
                                                    & 10                                                     & 1                                                       & 4.11      & \cellcolor[HTML]{EFEFEF}2.81      & 4.63                                  & \cellcolor[HTML]{EFEFEF}3.80                                & 7.38                                & \cellcolor[HTML]{EFEFEF}5.20                                \\
                                                    & 1                                                      & 10                                                      & 4.66      & \cellcolor[HTML]{EFEFEF}3.33      & 6.05                                  & \cellcolor[HTML]{EFEFEF}4.65                                & 4.49                                & \cellcolor[HTML]{EFEFEF}3.37                                \\
\multirow{-4}{*}{\textbf{Mean Squared Error}}       & 10                                                     & 10                                                      & 4.77      & \cellcolor[HTML]{EFEFEF}3.54      & 4.75                                  & \cellcolor[HTML]{EFEFEF}4.33                                & 5.58                                & \cellcolor[HTML]{EFEFEF}4.08                                \\ \hline
                                                    & 1                                                      & 1                                                       & 0.85      & \cellcolor[HTML]{EFEFEF}0.93      & 0.91                                  & \cellcolor[HTML]{EFEFEF}0.94                                & 0.86                                & \cellcolor[HTML]{EFEFEF}0.91                                \\
                                                    & 10                                                     & 1                                                       & 0.91      & \cellcolor[HTML]{EFEFEF}0.93      & 0.94                                  & \cellcolor[HTML]{EFEFEF}0.95                                & 0.84                                & \cellcolor[HTML]{EFEFEF}0.92                                \\
                                                    & 1                                                      & 10                                                      & 0.90      & \cellcolor[HTML]{EFEFEF}0.92      & 0.94                                  & \cellcolor[HTML]{EFEFEF}0.94                                & 0.91                                & \cellcolor[HTML]{EFEFEF}0.94                                \\
\multirow{-4}{*}{\textbf{Correlation Coefficient}}  & 10                                                     & 10                                                      & 0.89      & \cellcolor[HTML]{EFEFEF}0.92      & 0.95                                  & \cellcolor[HTML]{EFEFEF}0.95                                & 0.90                                & \cellcolor[HTML]{EFEFEF}0.93                                \\ \hline
                                                    & 1                                                      & 1                                                       & 0.5       & \cellcolor[HTML]{EFEFEF}0.9       & 0.4                                   & \cellcolor[HTML]{EFEFEF}0.7                                 & 0.4                                 & \cellcolor[HTML]{EFEFEF}0.6                                 \\
                                                    & 10                                                     & 1                                                       & 0.5       & \cellcolor[HTML]{EFEFEF}0.5       & 0.6                                   & \cellcolor[HTML]{EFEFEF}1.2                                 & 0.3                                 & \cellcolor[HTML]{EFEFEF}0.8                                 \\
                                                    & 1                                                      & 10                                                      & 0.4       & \cellcolor[HTML]{EFEFEF}0.8       & 0.5                                   & \cellcolor[HTML]{EFEFEF}0.6                                 & 0.6                                 & \cellcolor[HTML]{EFEFEF}0.7                                 \\
\multirow{-4}{*}{\textbf{Number of Crucial Errors}} & 10                                                     & 10                                                      & 1.2       & \cellcolor[HTML]{EFEFEF}1.3       & 1.1                                   & \cellcolor[HTML]{EFEFEF}1.2                                 & 0.5                                 & \cellcolor[HTML]{EFEFEF}0.3                                 \\ \hline
\end{tabular}
\label{overall_results}
\end{table} 
 
\subsection{Fuzzing an AES Implementation}
In order to provide first test results on real-world code we implemented AES on our DUT for data encryption. 
The AES algorithm does not perform input-dependent branches.
Hence, we can easily compare our calculated scores to the actual number of basic block transitions in our AES implementation.
Furthermore, AES is commonly performed by embedded devices, resulting in realistic power traces.
For our tests we encrypted randomly chosen plaintexts using one random key. 
During each encryption run, we let our implementation of side-channel aware fuzzing estimate the number of basic block transitions, which have been triggered at least once. 
Hence, we calculate the scores for each encryption and plaintext.
Our AES implementation triggers \num{41} unique basic block transitions during its ten rounds of encryption. 
We identified this number during a static code analysis of the compiled binary.
Over \num{100} encryption runs, our framework detects \num{38} transitions on average. 
This shows the applicability of our concept, regarding code commonly found on embedded devices.
In future work we will extend our tests to a broader range of software including more complex examples containing input-dependent branches.

\subsection{Transferability and Generalization} 
We carried out the description of the concept in a generic way. 
The motivation to do so is to give software engineers and testers the opportunity to adopt the concept, but implement it in a way fitting to their needs, attacked device, and measurement environment. 
These factors influence the quality of the power traces and hence the quality of the results. 
During the evaluation of different aspects of the concept and its implementation we payed attention to the transferability of side-channel aware fuzzing. 
Even though we fuzzed solely one specific DUT, a broader range of embedded devices can be analyzed with our concept. 
As can be seen in Section \ref{related_work}, considerable success has been achieved in side-channel based reverse engineering. 
In related work a significant part of the instruction sequence executed by different RISC-based target devices was successfully reconstructed.
With this state-of-the art research and the fact, that our concept is based on a successful branch detection we suffer no limitations regarding transferability to other RISC-based target devices.
In future work we will further investigate this assumption.

Furthermore we plan to target the challenge of detecting faults in embedded devices during fuzzing. 
Such faults often trigger a reboot of the device which is a known sequence of actions resulting in characteristic power traces. 
In an additional preprocessing step prior to the SCA calculations a machine learning based classifier could detect reboot sequences.
This information could then be sent to the fuzzing tool.

\section{Conclusion} \label{Conclusion}
In this paper we combine the two yet unlinked but well studied research fields of fuzzing and side-channel analysis to enable white-box fuzzing of software on embedded devices. 
With the results we gained from our proof of concept we show that the power side-channel provides sufficient information for a feedback-driven fuzzing loop. 

Side-channel aware fuzzing is a threefold concept in which we ultimately assign scores to fuzzing inputs.
The scores are proportional to the code coverage during the processing of the individual inputs. 
We use the number of basic block transitions to assess the code coverage and calculate the scores. 
In the machine learning based feature extraction approach we analyze the power consumption of the targeted embedded device. 
With the extracted features we discover the time when the device executes branch instructions which are the borders between basic blocks. 
Furthermore, we reconstruct the jump distances of the branches. 
In addition to that, we calculate fingerprints of the trace segments showing the basic blocks. 
We secondly conduct a control flow reconstruction using the extracted features. 
By evaluating either the branch distances or fingerprints of the trace parts between the branches, we are able to determine if we triggered a new basic block transition. 
In the final score calculation, we use the control flow of the software to calculate the score of the inputs sent to the target device. 

We carried out the proof of concept on a state-of-the-art ARM Cortex-M4 microcontroller. 
The structure of our synthetic test code allows calculation of the code coverage triggered by an input so that future work may employ it as a base line. 
Using this test code and an implementation of our concept, we are able to see a strong correlation between our calculated scores and the theoretical scores. 
The maximal correlation coefficient we achieved is \num{0.95}. 
Additionally we correctly estimated the number of basic block transitions in a light-weight AES implementation using our framework.
This states a significant step towards white-box fuzzing for vulnerability detection on embedded devices.

%
%
%
\bibliographystyle{splncs04}
\bibliography{mybib}

\end{document}